\documentclass[prb,twocolumn,showpacs,amsmath,amssymb]{revtex4-1}

\usepackage{graphicx}
\usepackage{multirow}
\usepackage{hyperref}
\hypersetup{backref=true,
 pdfnewwindow=true, colorlinks=true,
 linkcolor=blue, anchorcolor=blue,
 citecolor=blue, filecolor=blue,
 menucolor=blue, urlcolor=blue}
\usepackage{dcolumn}
\newcolumntype{.}{D{.}{.}{1}}

\begin{document}

\title{Proposal for a bulk material based on a monolayer FeSe on SrTiO$_3$ high-temperature superconductor}

\author{Sinisa Coh}
\email{sinisacoh@gmail.com} 
\author{Dung-Hai Lee}
\author{Steven G. Louie} 
\author{Marvin L. Cohen} 
\affiliation{Department of Physics, University of California, and
  Materials Sciences Division, Lawrence Berkeley National Laboratory,
  Berkeley, CA 94720, USA}

\date{\today}

\begin{abstract}
Motivated by the high superconducting transition temperature of monolayer FeSe on SrTiO$_3$, we propose a potential three-dimensional high-temperature superconductor superlattice FeSe--SrTiO$_3$ and study its structural stability and electronic structure using density functional theory. We find that the binding energy between the FeSe and SrTiO$_3$ layers is about $\sim$0.7~eV per (Fe$_2$Se$_2$) unit and that it saturates already within a single TiO$_2$ atomic layer of SrTiO$_3$.  In addition we analyzed the dynamical stability of the superlattice and compared it to the case of bulk SrTiO$_3$.
\end{abstract}

\pacs{74.78.Na, 74.20.Pq}

\maketitle

\section{Introduction}
\label{sec:introduction}

The discovery of superconductivity in single unit cell thick FeSe grown on the TiO$_2$-terminated surface of SrTiO$_3$ (1~UC~FeSe/STO)\cite{Xue} is remarkable in several respects.  Not only is the superconducting gap opening temperature  $T_{\rm gap}=55$--75~K estimated from angle-resolved photoemission spectroscopy\cite{XJZ1,XJZ2,DF1,DF2,XJZ3,JJ} (ARPES) the highest among all iron-based superconductors, it also has the simplest electronic structure. These findings have motivated many theoretical studies.\cite{coh2015,Lee,liu2012,bang2013,fawei2013}  Recently the mutual inductance measurement made on a sample showing a 65~K gap opening temperature in ARPES exhibited an onset of a Meissner effect at the same temperature.\cite{Yayu} This largely removed the long standing doubt of whether the energy gap measured by ARPES is caused by superconductivity.

In the bulk form, FeSe has a $T_{\rm c}$ of only 8~K at ambient pressure,\cite{hsu2008} and it peaks at 37~K under pressure.\cite{mizguchi2008,margadonna2009,medvedev2009}  This temperature is close to $T_{\rm c}\sim 30$~K found in $A_x$Fe$_{1-y}$Se at optimal doping with $A$ being either K, Rb, Cs, or Tl.\cite{DF3} A recently discovered bulk crystal consisting of FeSe layers intercalated with Li$_{1-x}$Fe$_x$OH again has a similar transition temperature of $\sim41$~K.\cite{xh,zhao} For potassium coated three monolayer FeSe on SrTiO$_3$ [K$_x$(FeSe)$_3$/STO] ARPES shows that the gap opens around 48~K.\cite{miyata,Wen,tang2015interface} All of the latter three superconductors have the same electronic structure as the 1~UC~FeSe/STO, but their  $T_{\rm c}$ is considerably lower.  This raises the question concerning the mechanism for the $T_{\rm c}$ enhancement in the 1~UC~FeSe/STO.  The fact that the measured Fermi surfaces of $A_x$Fe$_{1-y}$Se, (Li$_{1-x}$Fe$_x$OH)FeSe, and K$_x$(FeSe)$_3$/STO are nearly identical to that of 1~UC~FeSe/STO suggests that the reason for the enhanced  $T_{\rm c}$ is likely the close proximity of FeSe to SrTiO$_3$. 

It is suggested in Ref.~\onlinecite{JJ} by one of us and collaborators that the origin of the enhancement from 30--40~K to 55--75~K is the coupling between the FeSe electrons and SrTiO$_3$ phonons. On the other hand, the three other authors focused (Ref.~\onlinecite{coh2015}) on the role of the intrinsic coupling of the FeSe electrons to the FeSe phonons beyond the conventional density functional theory approach.  Now we briefly discuss the roles of SrTiO$_3$ phonons and FeSe phonons in these two studies.

According to Ref.~\onlinecite{JJ} the phonons in the SrTiO$_3$ enhance the intrinsic $T_{\rm c}$ from 30--40~K to 55--75~K (see Ref.~\onlinecite{Lee} for a review). Moreover, due to the small momentum transfer nature of the electron-phonon interaction, the coupling to the SrTiO$_3$ phonons enhances $T_{\rm c}$ regardless of the intrinsic pairing symmetry, as verified by a recent minus-sign-free quantum Monte Carlo simulation.\cite{Li} The evidence for a strong coupling between the FeSe electron and the SrTiO$_3$ phonon and its small momentum transfer nature is provided by the replication of all low-binding energy bands approximately 100~meV away, in the direction of higher binding energies. This replication was interpreted as a phonon shake-off effect\cite{H2,JJ} and is consistent with the presence of $\sim 100$~meV optical phonon band in SrTiO$_3$,\cite{phonon1,phonon2,phonon3} and similar replicas of the surface bands of the (001) surface of pure SrTiO$_3$.\cite{wang}

Regarding the contribution of  FeSe phonons to pairing, early local-density approximation calculations\cite{boeri2008,boeri2010} on related materials estimated this contribution to be too small to explain the experimentally found transition temperatures. On the other hand in Refs.~\onlinecite{mangal2014,coh2015} it is found that the strength of this interaction may have been severely underestimated in the early theoretical work.  The reason for this underestimation in earlier work is attributed in Ref.~\onlinecite{coh2015} to a tendency of a local-density approximation to underestimate the shearing (also called nematic, orthorhombic) instability relative to experiment in iron-based superconductors, as well as to strongly reduce the density of states at the Fermi level. In addition, scanning tunneling microscopy features\cite{tang2015interface} found in 1~UC~FeSe/STO are consistent with the calculated FeSe phonon spectral function\cite{coh2015} as well as with the kinks in the ARPES spectra on a similar material.\cite{Malaeb2014}    
   
In the present work we set aside the superconducting pairing mechanism intrinsic to FeSe and the validity of either of the above two suggested mechanisms.\cite{JJ,coh2015} Instead we focus on proposing bulk materials composed of stacked FeSe-SrTiO$_3$ interfaces for possible further enhancement of $T_{\rm c}$. Thus our purpose here is to study the structural stability and electronic structure of a novel bulk material, first proposed in Ref.~\onlinecite{Lee}, in which FeSe layers are bonded from both sides by TiO$_2$ terminated layers of SrTiO$_3$. Our motivation is to double the interface between FeSe and SrTiO$_3$, relative to the case of 1~UC~FeSe/STO.  Due to the exponential sensitivity of $T_{\rm c}$ to the pairing strength, this may lead to a even larger $T_{\rm c}$ enhancement.\cite{JJ,Lee} In addition, the three-dimensionality of the proposed material should suppress the superconducting phase fluctuation in the two-dimensional 1~UC~FeSe/STO (such phase fluctuation is observed in Ref.~\onlinecite{Yayu}).

\section{Methods and results}
\label{sec:results}
We now discuss the results of our density functional theory (DFT) based first-principles calculations of the FeSe-SrTiO$_3$ superlattices performed using the Quantum-ESPRESSO package.\cite{giannozzi}  Most of our calculations use the Perdew-Burke-Ernzerhof\cite{pbe1997} (PBE) exchange-correlation functional.  However, for a more accurate determination of the binding energy we use the vdW-DF2\cite{vdwdf2} functional that includes van der Waals interactions.  The ultrasoft pseudopotentials used are from the GBRV\cite{Garrity2014} library with 40 and 200~Ry kinetic energy cutoffs for the electron wavefunction and the charge density.  The Gaussian smearing is set to 13~meV and a 14$^3$ $k$-point grid is used.  To correct the band gap of SrTiO$_3$ we apply Hubbard +$U$ correction on the Ti atom with $U=6$~eV, which gives a $\Gamma$ point energy gap of 3.0~eV in bulk SrTiO$_3$. Electron doping is compensated with a uniform positive background, to keep the computational unit cell neutral.  All structural relaxation is done without electron doping.

It is important to note that certain features of the ARPES measured band structure of 1~UC~FeSe/STO and related FeSe based superconductors are not well reproduced by a conventional DFT calculation.  Some aspects of the measured band structure are better captured by DFT calculation assuming a nonmagnetic (NM) ground state while others agree with assuming a checkerboard antiferromagnetic (c-AFM) ground state. In particular, ARPES finds no hole pockets at the zone center and two pockets at the $M$ point.  In DFT there are no hole pockets at the zone center in the c-AFM case (and not in NM) while there are two pockets at the M point in the NM case (and not in c-AFM).  However, structural stability, the main focus of this work, is largely the same in the NM and the c-AFM states as we demonstrate later in this paper.

\subsection{Structure}

\begin{figure}[!t]
\centering
\includegraphics{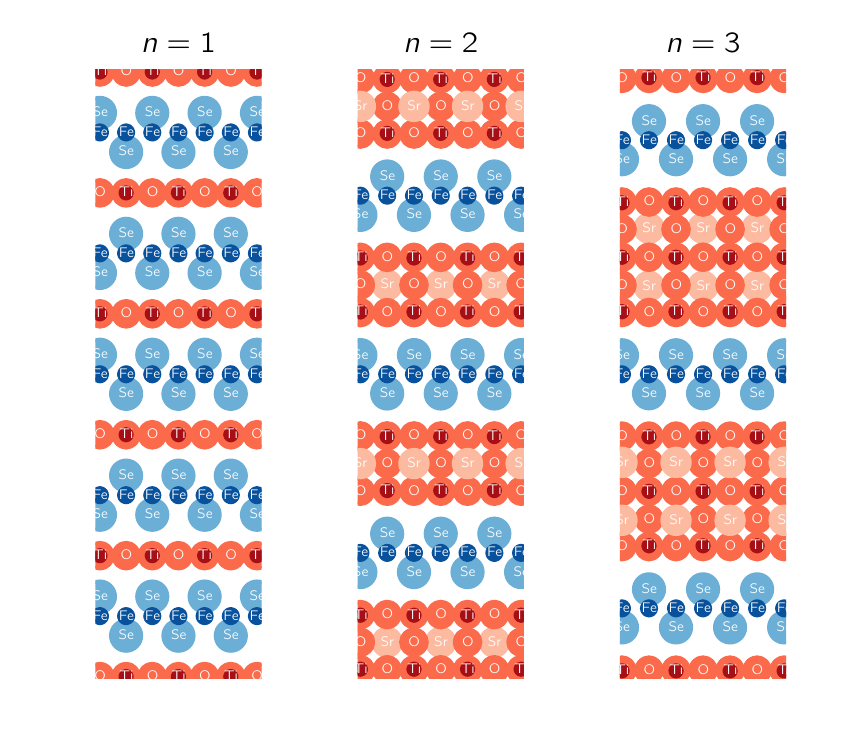}
\caption{\label{fig:structure} Three FeSe-SrTiO$_3$ superlattices ($n=1,2,3$) studied in this work with varying number of layers between FeSe.  Dark and light blue circles correspond to Fe and Se atoms while dark, medium, and light red circles correspond to Ti, O, and Sr atoms.}
\end{figure}

We focus here on structures where FeSe is on both sides interfaced with TiO$_2$ terminated surfaces of SrTiO$_3$, both because that is a significantly lower energy interface (0.65~eV versus 0.44~eV per Fe$_2$Se$_2$ formula unit for $n=3$, see Table~\ref{tab:struc}) and because high $T_{\rm c}$ in the monolayer FeSe on SrTiO$_3$ was observed for this type of interface. While bulk SrTiO$_3$ contains equal numbers of TiO$_2$ and SrO layers, our proposed superlattice contains exactly one more TiO$_2$ layer than the number of SrO layers. The repeat unit of our superlattice is therefore
\begin{center}
---Fe$_2$Se$_2$---TiO$_2$---$\Big($SrO---TiO$_2$---$\Big)_{n-1}$
\end{center}
for any positive integer $n$. Figure~\ref{fig:structure} shows the $n=1$, 2, and 3 superlattices we studied. The structure proposed in Ref.~\onlinecite{Lee} is the one indexed by $n=2$.  Table~\ref{tab:struc} shows the binding energies and other structural parameters.  We computed the binding energy by comparing the energy of the entire superlattice with that of isolated FeSe and isolated SrTiO$_3$ slab with the same number of layers and the same in-plane lattice constants. 

\subsubsection{Binding energy}

\begin{table*}[!t]
  \caption{\label{tab:struc} This table contains the binding energies of FeSe-SrTiO$_3$ superlattices per one formula unit (Fe$_2$Se$_2$),  the relaxed in-plane lattice constant ($a$), the selenium height relative to the plane of iron atoms, the distance from Fe in FeSe to Ti in the top-most TiO$_2$ layer, the rumpling in the Ti-O surface, and the magnetic moment ($\mu$) per iron atom (in the c-AFM state).  The binding energy is given both with and without van der Waals interaction in the nonmagnetic (NM) and the checkerboard antiferromagnetic (c-AFM) states. The inference on the effects of magnetic states on the binding energies is made by comparing the second and third columns. The binding energy quoted in the text are computed with the van der Waals interaction in the nonmagnetic state. The remaining quantities are calculated without van der Waals correction in the c-AFM state.}  
\begin{ruledtabular} 
\begin{tabular}{lcccccccc}
 & \multicolumn{3}{c}{Binding energy per Fe$_2$Se$_2$} & $a$ & Se height  & Fe-Ti & Ti-O rumpling & $\mu$ \\
 \cline{2-4}
 & with vdW &  \multicolumn{2}{c}{\it without vdW} \\
  \cline{2-2} \cline{3-4}
 & NM & {\it NM} & {\it c-AFM} \\
 & (eV) & {\it (eV)} & {\it (eV)}  & (\AA) & (\AA) & (\AA) & (\AA) & ($\mu_{\rm B}$)\\ 
\hline
\multicolumn{2}{l}{FeSe-SrTiO$_3$ superlattices}\\
\quad \quad $n=1$
& 0.79 & {\it 0.36} & {\it 0.41} & 3.78 & 1.41 & 4.37 &       & 2.14 \\
\quad \quad $n=2$
& 0.65 & {\it 0.17} & {\it 0.20} & 3.87 & 1.39 & 4.50 & 0.046 & 2.33 \\
\quad \quad $n=3$
& 0.65 & {\it 0.15} & {\it 0.18} & 3.90 & 1.37 & 4.52 & 0.054 & 2.41 \\ \\
\multicolumn{2}{l}{Reference points}\\
\quad \quad Relaxed FeSe monolayer             
&      &   &   & 3.82 & 1.41 &      &       & 2.20 \\
\quad \quad SrTiO$_3$                          
&      &   &   & 3.97 &      &      & 0.073 &      \\ \\
\multicolumn{2}{l}{Alternative configurations}\\
\quad \quad SrO termination (n=3)          
& 0.44 &    &  &      &      &      &       &      \\
\quad \quad primitive tetragonal (n=3)         
& 0.61 &    &  &      &      &      &       &      \\
 \end{tabular}
 \end{ruledtabular}
\end{table*}

As shown in Table~\ref{tab:struc} the binding energy of the superlattice per one Fe$_2$Se$_2$ formula unit is nearly independent of $n$ for $n>1$.  We find it to equal 0.79~eV for the thinnest superlattice ($n=1$) and 0.65~eV for two thicker cases ($n=2,3$).  Therefore we conclude that the cohesive energy between the FeSe and SrTiO$_3$ components of the superlattice is reached already within a single TiO$_2$ layer.  A similar trend is found for the structural parameters such as the in-plane lattice constant, the selenium height, and the distance between FeSe and TiO$_2$ layers (see Table~\ref{tab:struc}).

In the lowest energy structure, Se atoms in the FeSe layer are directly above/below the Ti atoms in the TiO$_2$ layer. For this reason, the lowest energy structure is in the body centered tetragonal phase (see Fig.~\ref{fig:structure}).  In the primitive tetragonal phase, where Se is above/below Ti on only one side of FeSe, the binding energy per Fe$_2$Se$_2$ unit is reduced from 0.65~eV to 0.61~eV (in the $n=3$ case).

Comparing the relaxed in-plane lattice constant of the superlattice, we find that it is larger than that of isolated FeSe monolayer and smaller than that of bulk SrTiO$_3$ (Table~\ref{tab:struc}), with the exception of $n=1$ case.  As expected, when $n$ increases the lattice constant of the superlattice approaches that of bulk SrTiO$_3$.

We find nearly no effect of the magnetic ground state on the cohesive energy of the superlattice.  As shown in the second and third columns of Table~\ref{tab:struc} the  binding energy without the van der Waals interaction of the superlattice in the NM and c-AFM state differ only between 30 and 50~meV.

\subsubsection{Polar instability}

\begin{table*}[!t]
  \caption{Phonon instabilities at high-symmetry $\bf q$ points in the bulk SrTiO$_3$ and $n=3$ superlattice. We divide instabilities into two classes: those due to a polar distortion and those due to oxygen octahedra rotations.
   \label{tab:dyn}
  }  \begin{ruledtabular} \begin{tabular}{lcccl}
&  \multicolumn{1}{c}{Frequency}   & Degeneracy & $\bf q$ &  Type of instability \\
&  \multicolumn{1}{c}{(cm$^{-1}$)} &      &         &     \\ 
\hline
Polar Ti-O distortion\\
\quad \quad  Bulk SrTiO$_3$\\
& $140i$ & $3$ & $(0 \ 0 \ 0)$ &  Polar, Ti-O\\
& $50i$  & $2$ & $(\frac{1}{2} \ 0 \ 0)$ &  Antipolar, Ti-O\\
\multicolumn{4}{l}{\quad \quad The $n=3$ superlattice}\\
& $122i$ & $2$ & $(0 \ 0 \ 0)$ & In-plane polar, Ti-O\\
&  $78i$ & $2$ & $(0 \ 0 \ 0)$ & In-plane polar, mostly O\\
&  $62i$ & $2$ & $(0 \ 0 \ 0)$ & In-plane anti-polar,\footnotemark[1] Ti-O \\
&  $51i$ & $2$ & $(0 \ 0 \ 0)$ & In-plane polar, mostly O \\
\\
Oxygen octahedron rotation\\
\quad \quad  Bulk SrTiO$_3$\\
&  $7i$  & $1$ & $(0 \ \frac{1}{2} \ \frac{1}{2})$ &  In-phase oxygen octahedra rotation \\
& $75i$  & $3$ & $(\frac{1}{2} \ \frac{1}{2} \ \frac{1}{2})$ &  Out-of-phase oxygen octahedra rotation \\
\multicolumn{4}{l}{\quad \quad The $n=3$ superlattice}\\
& $122i$ & $1$ & $(\frac{1}{2} \ \frac{1}{2} \ 0)$ &  Rotation of middle oxygen octahedron, along $\hat{z}$ \\
&  $60i$ & $2$ & $(\frac{1}{2} \ \frac{1}{2} \ 0)$ &  In-phase oxygen octahedra rotation around $\hat{x},\hat{y}$ \\
 \end{tabular}
 \end{ruledtabular}
     \footnotetext[1]{In the $n=3$ superlattice primitive unit cell contains three Ti atoms.  Therefore this mode is antipolar despite the fact that ${\bf q}=(0 \ 0 \ 0)$.  In this particular mode Ti atoms at the opposite end of the superlattice move in opposite directions.}
\end{table*}

Bulk SrTiO$_3$ is known to be on the verge of a polar instability. Its dielectric constant is nearly divergent\cite{sto1962} at low temperatures (below 50~K).  Density functional theory calculations without quantum fluctuations\cite{zhong1995, sai2000} find unstable phonon modes at the zone center corresponding to polar distortion.  These modes are stabilized only by the inclusion of quantum fluctuation effects.\cite{zhong1996}

However, the surface of SrTiO$_3$, unlike the bulk, is known to be polar\cite{surf1989} due to a slight displacement (rumpling) of oxygen atoms relative to titanium atoms.  The direction of the atomic displacement is such that the oxygen atoms are moved away from the TiO$_2$ plane towards the vacuum region. Our calculations show this rumpling to be equal 0.073~\AA~ on a pristine SrTiO$_3$ surface (see Table~\ref{tab:struc}) while it is somewhat reduced in the FeSe-SrTiO$_3$ superlattices to 0.046 for $n=2$ and 0.054~\AA\ for $n=3$.  The direction of the rumpling is the same as for the pristine SrTiO$_3$ surface.  Since the magnitude of the rumpling distortion is so similar, we conclude that surface polar instabilities of SrTiO$_3$ are likely not affected much by the presence of the FeSe layer in the superlattice.

To further compare polar instabilities in bulk SrTiO$_3$ and the superlattice, we computed phonon frequencies in pure SrTiO$_3$ and the $n=3$ superlattice at high-symmetry $\bf q$ vectors.  These frequencies are reported in Table~\ref{tab:dyn} and they do not include the LO-TO correction at ${\bf q}=0$ (i.e., in all cases we report transverse optical, TO, modes only). For phonons associated with the structure instabilities, the frequencies are purely imaginary.

In bulk high-symmetry cubic SrTiO$_3$ we find several unstable phonon modes. For example, at the Brillouin zone center ${\bf q}= (0\ 0\ 0)$ we find a phonon triplet with a frequency of $140i$~cm$^{-1}$ corresponding to the polar modes in which Ti and O atoms move in opposite directions. At the three equivalent centers of the Brillouin zone faces such as ${\bf q}= (\frac{1}{2}\ 0\ 0)$ we find a doublet of unstable modes with a frequency of $50i$~cm$^{-1}$ corresponding to the antipolar distortion of Ti and O atoms.

In the $n=3$ FeSe-SrTiO$_3$ superlattice we find very similar frequencies of unstable polar phonon modes ranging from $122i$~cm$^{-1}$ to $51i$~cm$^{-1}$, as shown in Table~\ref{tab:dyn}.  The out-of plane polar mode is stabilized in the superlattice (its frequency is $330$~cm$^{-1}$) since there is a static polar distortion (the Ti-O rumpling) at the FeSe-SrTiO$_3$ interface.  The antipolar mode in the superlattice has nearly the same frequency as in the bulk (it is $62i$~cm$^{-1}$ versus $50i$~cm$^{-1}$ in the bulk).  In the superlattice case the antipolar mode appears at the zone center ${\bf q}=(0 \ 0 \ 0)$ since the primitive unit cell of the superlattice contains three Ti atoms.

\subsubsection{Rotation of oxygen octahedra}

We now turn to the analysis of the remaining unstable modes corresponding to the rotation of oxygen octahedra.  Unlike polar and antipolar distortions, oxygen octahedral distortions are not removed by quantum fluctuations and static distortion occurs in bulk SrTiO$_3$ at temperatures below 105~K.

In cubic bulk SrTiO$_3$ we find unstable modes at the $(0 \ \frac{1}{2} \ \frac{1}{2})$ point corresponding to the in-phase rotations with a frequency of $7i$~cm$^{-1}$.  On the other hand, out-of-phase octahedral rotations are even more unstable since their frequency is $75i$~cm$^{-1}$. Here by in-phase rotation we have in mind a rotation in which adjacent octahedra along the rotation axis rotate in the same sense.  This kind of rotation is also denoted as a "$+$"~rotation by Glazer.\cite{Glazer:a09401}  The out-of-phase rotations are denoted as a "$-$"~rotation.

In the case of the superlattice the strongest rotational instability has a frequency of $122i$~cm$^{-1}$ and it corresponds to the rotation of the oxygen octahedron in the middle of the SrTiO$_3$ slab around the $z$ axis (the $z$ axis is perpendicular to the slab). However, the calculated condensation energy of this mode is only 75~meV per primitive unit cell of $n=3$ superlattice.  Another instability with a frequency of $60i$~cm$^{-1}$ corresponds to the in-phase rotation of octahedra around the $x$ and $y$ axes.

\begin{figure*}[!t]
\centering
\includegraphics{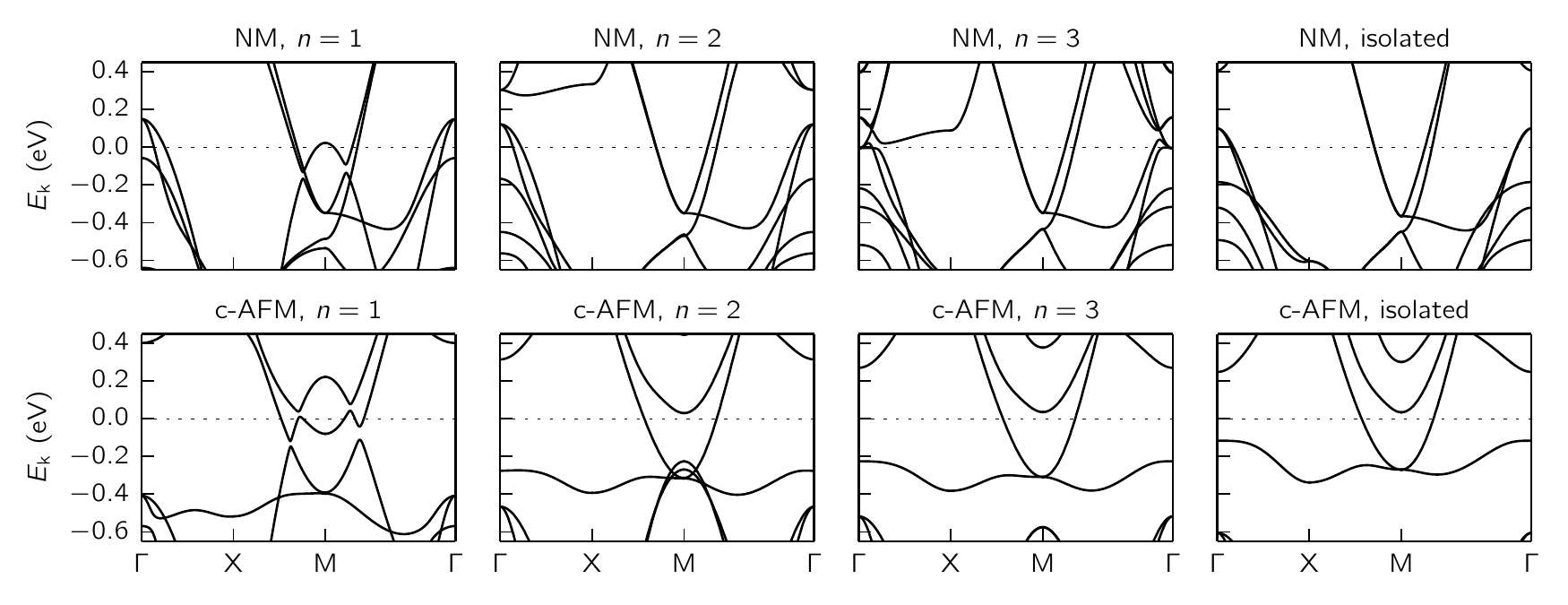}
\caption{\label{fig:bs} 
Comparison of band structures with varying layer index $n$ and a fixed doping level of 0.12~electrons per single Fe atom.  The right-most panel shows the band structure of an isolated FeSe monolayer with a lattice constant of bulk SrTiO$_3$ for comparison.  The top row panels show the band structure in the nonmagnetic (NM) case while the bottom row panels correspond to the checkerboard antiferromagnetic (c-AFM) case.}
\end{figure*}

Therefore, we can conclude that all structural instabilities in the superlattice are originating from the SrTiO$_3$ layers and are present even in bulk SrTiO$_3$.  While polar distortions are slightly suppressed in the superlattice, oxygen octahedral distortions are somewhat enhanced (from $75i$~cm$^{-1}$ to $122i$~cm$^{-1}$).  Therefore, we expect that the same series of structural phase transitions will appear in the superlattice as in the SrTiO$_3$ slab. 

\subsection{Electronic structure}

We now turn to the electronic structure of the superlattices.   As in the case of 1~UC~FeSe/STO, we expect that the FeSe layers in our superlattice will be electron-doped by oxygen vacancies in the SrTiO$_3$ layers. In fact, oxygen vacancies are a common occurrence in SrTiO$_3$ and other perovskite oxides.  To simulate the effect of oxygen vacancies in our calculations, we added in the calculation excess electrons along with a uniform positive charge background, to keep the computational unit cell charge-neutral.  The concentration of the added electron density is 0.12 electrons per single Fe atom, the same as estimated from the Fermi volume measured by ARPES in 1~UC~FeSe/STO.

The calculated band structures are shown in Fig.~\ref{fig:bs} for the $n=1,2,3$ superlattices.  For a comparison, we also show a band structure of an isolated FeSe monolayer but with an lattice constant equal to that of bulk SrTiO$_3$.  In all four cases, we show band structures in the NM and the c-AFM states for a comparison.  The band structure near the Fermi level of all superlattices we studied are very two-dimensional.  For example we find that the $z$ axis dispersion at the $M$ point is at most 2~meV.

As shown in the left-most panel of Fig.~\ref{fig:bs}, in the case of the $n=1$ there is a hole pocket at $M$ originating from oxygen $p$ states at the Fermi level (even with included electron doping of 0.12~electrons per Fe).  However, as $n$ is increased, the top of the oxygen band moves below the Fermi level. The energy separation between the top of the oxygen band and the Fermi level is roughly linearly proportional to $n$.  Already at $n=2$ we find that the Fermi level electronic structure is dominated by FeSe and is nearly indistinguishable from the case of an isolated FeSe monolayer (rightmost panel in Fig.~\ref{fig:bs}).

\section{Summary and outlook}
\label{sec:final}

The single layer of FeSe or FeAs is a common structural motif for all iron-based superconductors.  These FeSe or FeAs layers are commonly thought to be where conduction and superconductivity occurs while the in-between buffer layers serve only as reservoirs of charge and for structure stabilization.  Here we are proposing a family of superlattices in which the in-between layers serve an active role in the superconductivity by enhancing the superconducting transition temperature $T_{\rm c}$ of the material.

We expect a high superconducting transition temperature in FeSe-SrTiO$_3$ superlattices for the following three reasons.  First, SrTiO$_3$ layers adjacent to the FeSe layers can electron dope them and thus allow the intrinsic pairing mechanism operating in heavily doped FeSe to act.  Second, the enhancement mechanism due to SrTiO$_3$ phonons proposed in Ref.~\onlinecite{JJ} is doubled in the superlattice.  Third, three-dimensionality of the superlattice will suppress the superconducting phase fluctuations and thus enhance $T_{\rm c}$.  Indeed in Ref.~\onlinecite{Yayu} it is shown that in the two-dimensional case (1 UC FeSe/STO) the full diamagnetism is not achieved until $\sim$15~K despite the fact that the Meissner effect onset occurs at 65~K.  This wide phase transition is consistent with suppression of superconductivity by phase fluctuations in two dimensions.

Furthermore, we expect the $T_{\rm c}$ enhancement mechanism due to the substrate phonons proposed in Ref.~\onlinecite{JJ} to hold for other oxides. For example earlier work found that one can replace SrTiO$_3$ with BaTiO$_3$ and have a similar $T_{\rm c}$ enhancement\cite{DF2} as in 1~UC~FeSe/STO.   This observation is consistent with the fact that BaTiO$_3$ has similar high energy phonon bands as SrTiO$_3$. In addition, we expect that the structural-template effect\cite{coh2015} on a cubic SrTiO$_3$, as well as doping due to oxygen vacancies, will continue to hold for other transition metal oxides as well.

We hope that FeSe-SrTiO$_3$ superlattices might be grown by bulk crystal growth techniques molecular beam epitaxy (MBE), or pulsed layer deposition (PLD) methods. For the latter two techniques controlling the chemical potential of all five elements in the superlattice (Fe, Se, Sr, Ti, and O) and the different growth temperatures for FeSe and SrTiO$_3$ poses strong challenges. On the other hand, we expect the $T_{\rm c}$ enhancement mechanism to hold for even binary oxides such as TiO$_2$ (our $n=1$ superlattice), which might be grown more easily with the MBE or PLD method. In fact, in a recent study\cite{tio2} an FeSe monolayer was successfully grown on top of TiO$_2$ anatase.

\begin{acknowledgments}
This work was supported by the Theory Program at the Lawrence Berkeley National Lab through the Office of Basic Energy Sciences, U.S. Department of Energy under Contract No. DE-AC02-05CH11231 which provided for the theoretical analysis and the National Science Foundation under Grant No. DMR15-1508412 which provided for the structural calculation. This research used resources of the National Energy Research Scientific Computing Center, which is supported by the Office of Science of the U.S. Department of Energy.
\end{acknowledgments}

\bibliography{pap}

\end{document}